# Constraints on color-flavor locked quark matter in view of the HESS J1731-347 measurement

K. Kourmpetis[1,2,*], P. Laskos-Patkos[1], Ch. C. Moustakidis[1]

[1] *Department of Theoretical Physics, Aristotle University of Thessaloniki, 54124 Thessaloniki, Greece*
[2] *Department of Physics, Eberhard Karl University of Tübingen, 72074 Tübingen, Germany*

**Abstract**   Astrophysical observations play a crucial role in understanding the processes within compact stars. A recent study measured the central object in the HESS J1731-347 supernova remnant (SNR), estimating its mass at $M = 0.77^{+0.20}_{-0.17} M_\odot$ and radius at $R = 10.40^{+0.86}_{-0.78}$ km, identifying it as the lightest neutron star ever observed. Conventional models suggest neutron stars form with a minimum gravitational mass of approximately $1.17 M_\odot$, raising the question of whether this object is a typical neutron star or possibly an "exotic" star. To investigate, we utilize the Color-Flavor Locked (CFL) equation of state (EoS), integrating data from the HESS J1731-347 measurement with pulsar observations and gravitational wave detections. Additionally, we construct hybrid EoS by combining the MDI-APR1 (hadronic) and CFL (quark) EoS, introducing a phase transition through Maxwell construction. Our findings reveal that absolutely stable CFL quark matter effectively explains all observed measurements, including the central object of HESS J1731-347, whereas hybrid models incorporating the CFL MIT Bag model cannot account for the masses of the most massive observed pulsars.

## INTRODUCTION

The study of the equation of state of nuclear matter plays a pivotal role in understanding the fundamental properties of compact stars. These exotic objects provide a unique laboratory for probing the behavior of matter under intense conditions [1-4]. The recent HESS J1731-347 measurement [5], a central compact object (CCO) [6-8] with an unusually low mass, has sparked renewed interest in exploring alternative forms of matter beyond traditional neutron stars. This intriguing discovery challenges our current understanding and opens the door to new possibilities in the study of dense matter physics.

The nature of the HESS J1731-347 CCO poses a strong theoretical challenge as up to this moment it is not clear how such light neutron stars could be produced in supernova explosions [9]. The peculiar characteristics of this compact object (mass and radius) may suggest that neutron matter alone may not be sufficient to explain the observed properties, necessitating the consideration of more exotic forms of matter. Quark stars and hybrid stars represent viable alternatives that must be investigated to fully understand the implications of this object. Brodie and Haber [10] further emphasized that pure nuclear matter alone may be insufficient to account for the HESS J1731-347 CCO, whereas quark and hybrid models may provide more accurate predictions [10-12].

In this work, we follow the proposal by Di Clemente et al. [11] and Horvath et al. [12], considering the CCO in HESS J1731-347 to be a quark strange star (SS) [13-18]. Strange quark matter (SQM) is considered a strong candidate for the true ground state of matter [19–23]. Interestingly, the recent study of Ref. [24] has posed tight constraints on SQM's possible existence. Future works may finally provide a conclusive answer on the nature of the absolutely stable state of matter.

As SQM, we employ the Color-Flavor Locked (CFL) model. At asymptotically high densities, quark matter will enter a color-superconducting phase known as the Color-Flavor Locked (CFL) state [25-28], where quarks of different colors and flavors form Cooper pairs [29, 30]. The CFL matter, also suggested in Ref. [31], could potentially explain the nature of the object observed in the HESS J1731-347 SNR. Astrophysical constraints on color-superconducting phases have also been explored in Ref.

---

\* Corresponding author: kkourmpe@physics.auth.gr

[32].

In scenarios where the quark phase is not absolutely stable but emerges as the dominant phase at sufficiently high densities, hybrid stars are formed [33-37]. In the present work, we construct a hybrid model that combines the MDI-APR1 EoS [38] with CFL matter, featuring a Maxwell phase transition [2]. Hybrid matter is also considered a strong candidate for the nature of the HESS J1731-347 CCO [10, 39-45].

Any proposed EoS must explain the HESS J1731-347 measurement while aligning with observations of massive pulsars [46-48] and the GW170817 merger [49]. It should universally describe compact star physics and meet compactness constraints, including the black hole limit [50], Buchdahl limit [51], and causality limit [52].

The motivation behind this study is to impose advanced and precise constraints on the bag constant and superconducting gap parameter space within the CFL EoS. By exploring various combinations of these parameters, we aim to identify the specific value pairs that not only predict the existence of the HESS J1731-347 CCO, but also accurately describe the properties of the GW170817 event and the most massive known pulsars. Additionally, we investigate whether a CFL hybrid model can provide a viable explanation for the properties of all these objects. The findings will contribute to our understanding of the dense matter EoS and may offer new insights into the possible existence of quark and hybrid stars.

This paper is organized as follows: in section two, we present the theoretical model utilized in this study. Section three provides the results for each model and a detailed discussion of these findings. Finally, in section four, we summarize the key insights and conclusions drawn from this work.

**THE MODEL**

*TOV Equations*

The structure of neutron stars is encapsulated in the so-called TOV equations, named after Tolman, Oppenheimer and Volkov [53]. The TOV equations are given by [54]:

$$\frac{dP(r)}{dr} = -\frac{GM(r)\rho(r)}{r^2}\left(1 + \frac{P(r)}{\rho(r)c^2}\right)\left(1 + \frac{4\pi r^3 P(r)}{c^2 M(r)}\right)\left(1 - \frac{2GM(r)}{c^2 r}\right)^{-1} \quad (1)$$

$$\frac{dM(r)}{dr} = 4\pi r^2 \rho(r), \quad (2)$$

where ρ(r) is the matter density and ε(r) is the energy density. The system is solved after transforming the TOV equations into a form that is more suitable for numerical integration [55].

*Color-Flavor-Locked Equation of State*

The EoS for CFL quark matter can be derived in the MIT bag model framework [56]. To the order of $\Delta^2$ and $m_s^2$ ($m_s$ represents the mass of the strange quark, μ the quark chemical potential and B the bag constant) the pressure and energy density can be expressed as follows ($\hbar = c = 1$) [28]:

$$P = \frac{3\mu^4}{4\pi^2} + \frac{9\alpha\mu^2}{2\pi^2} - B, \quad \text{and} \quad \varepsilon = \frac{9\mu^4}{4\pi^2} + \frac{9\alpha\mu^2}{2\pi^2} + B, \quad (3)$$

$$\text{where} \quad \alpha = -\frac{m_s^2}{6} + \frac{2\Delta^2}{3} \quad (4)$$

Combining the above equations one can obtain an analytic expression for ε(P) and P(ε):

$$\varepsilon = 3P + 4B - \frac{9\alpha\mu^2}{\pi^2}, \quad \text{where} \quad \mu^2 = -3\alpha + \left[\frac{4}{3}\pi^2(B+P) + 9\alpha^2\right]^{1/2} \quad (5)$$

$$P = \frac{\varepsilon}{3} - \frac{4B}{3} + \frac{3\alpha\mu^2}{\pi^2}, \quad \text{where} \quad \mu^2 = -\alpha + \left[\alpha^2 + \frac{4}{9}\pi^2(\varepsilon - B)\right]^{1/2} \quad (6)$$

For CFL quark matter to be absolutely stable, its energy per baryon must be lower than the neutron mass ($m_n$) at zero pressure (P = 0) and temperature (T = 0) [18]. Consequently [28]:

$$\left.\frac{\varepsilon}{n_B}\right|_{P=0} = 3\mu \leq m_n = 939 \text{ MeV} \qquad (7)$$

This result directly follows from the shared Fermi momentum among the three quark flavors in CFL matter, valid at T = 0 without any approximation. Since this condition must be satisfied at zero pressure, using Eq. (5), we have [57]:

$$B < -\frac{m_s^2 m_n^2}{12\pi^2} + \frac{\Delta^2 m_n^2}{3\pi^2} + \frac{m_n^4}{108\pi^2} \qquad (8)$$

This equation defines a region in the $m_s$-B plane where the energy per baryon is less than $m_n$ for a specified value of $\Delta$ [57]. This condition combined with the constraint that two-flavored quark matter should be less stable than nuclear matter, or B ≥ 57 MeV·fm$^{-3}$ [18] in the MIT Bag Model framework, describes the so-called stability windows [57, 58].

*Phase Transition: Maxwell Construction*

In this work on hybrid stars, we combine neutron matter models with quark matter in the CFL phase to develop hybrid star equations of state. For the neutron phase, we adopt the MDI-APR1 EoS. The hadron to quark phase transition is treated as a first-order phase transition using the Maxwell construction. Thermodynamic equilibrium between the two phases-phase I (hadronic) and phase II (quark)-is established when [2] (thermal equilibrium is trivially satisfied due to the use of T = 0 EoS):

$$P_I(\mu_B^I) = P_{II}(\mu_B^{II}), \quad \text{with} \quad \mu_B^I = \mu_B^{II} \qquad (9)$$

where P is the pressure and $\mu_B$ is the baryon chemical potential of each phase.

## RESULTS AND DISCUSSION

*Pure CFL Quark Matter*

In our analysis, we employ the expression that defines the CFL stability windows, see Eq. (8), while maintaining the mass of the strange quark constant at $m_s$ = 95 MeV. This function is plotted in the B-$\Delta$ space combined with the bag constant constraint resulting in a new stability window, as illustrated in Fig. 1 (left). The CFL equations of state used in this work are included in this figure and they all reside within the stability window. For the CFL models utilized, we have used typical B and $\Delta$ values (B typical values are ~57-150 MeV·fm$^{-3}$ and $\Delta$ values are ~50-150 MeV).

Having examined the stability of the CFL matter, we proceed by constructing the M-R diagram for the CFL EoS models employed. This graph is shown in Fig. 1 (right). We focus on the curves that satisfy all the criteria, including the one imposed by the heaviest pulsar, PSR J0952-0607. It is evident that the EoS that meet all the requirements are CFL-2, CFL-3 and CFL-6. CFL-2 and CFL-6 both have a maximum mass in the mass-range of PSR J0952-0607 and share an analogous relationship between B and $\Delta$ values, with B being 50-60% of $\Delta$, while CFL-3 exceeds this range, having B = 40% of $\Delta$. We find that if B is lower than 50% of $\Delta$, the maximum mass would potentially exceed the mass-range of PSR J0952-0607. This is clearly illustrated in Figure 2 (left), where we have plotted the M-R diagram for B values of 50% $\Delta$ and 60% $\Delta$, showing that all the curves pass through all the constraints imposed by observations, while peaking inside the PSR J0952-0607 mass-range.

We continue our analysis by generating M-R curves for a range of B and $\Delta$ values, assigning them typical values, and identifying which of them align with all measurements. This is illustrated in Fig. 2 (right). The area of interest is the greenish region, where all three events are predicted. The graph also accounts for the minimum value of the bag constant, constraining the B-$\Delta$ pairs inside a triangle (considering the aforementioned typical values). This more accurate result agrees with the B and $\Delta$ analogy constraints we roughly estimated in our previous analysis, as seen from the orange area of Fig. 2 (right). We note that CFL matter adheres to the causality limit, with sound speed consistently below $c_s = 1$ and asymptotically approaching the conformal limit of $c_s^2 = 1/3$, as also noted by Flores and Lugones [57].

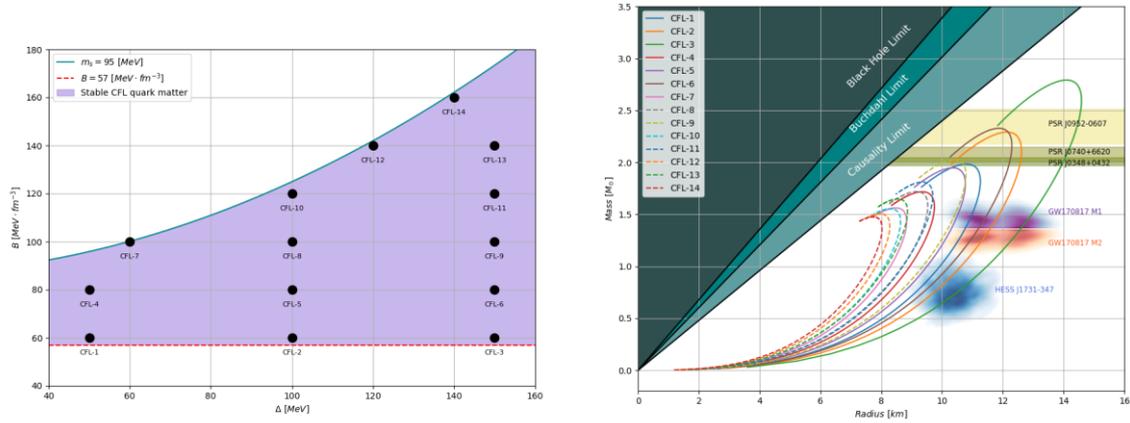

**Figure 1**. *CFL stability window for $m_s = 95$ MeV, including the CFL EoS employed. The red dashed line is the minimum value of B (B ≥ 57). The colored area indicates the range of B and Δ, for the given $m_s$ value, where the CFL EoS is stable (left panel). M-R diagram for the CFL EoS presented in the CFL stability window. The graph includes constraints forced by pulsar observations (PSR J0348+0432, PSR J0740+6620, PSR J0952-0607), the GW170817 merging event and the CCO in the HESS J1731- 347 SNR [5, 46-49] (right panel).*

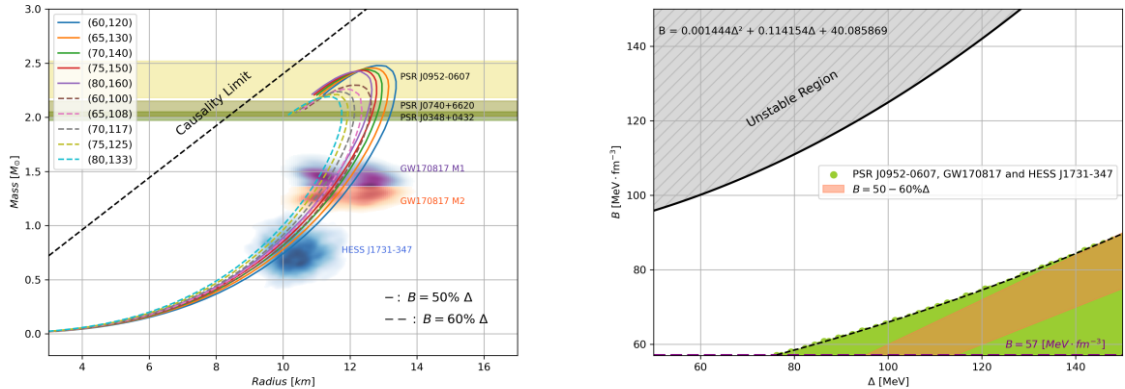

**Figure 2**. *M-R diagram that aligns with measurements [5, 46-49]. The solid line represents the EoS with B = 0.5Δ, while the dashed ones the EoS with B = 0.6Δ. The legend is in the form of (B in MeV·fm$^{-3}$, Δ in MeV) (left panel). Constraints in the B-Δ plane. The greenish area, formed by the upper boundary fit and the $B_{min}$ (purple), represents B-Δ pair combinations that produce M-R curves that align with the PSR J0952-0607, GW170817 and HESS J1731-347 SNR observation. The shaded region marks the unstable area (right panel).*

*Hybrid CFL Matter*

In Fig. 3, we present the mass-radius diagrams for the hybrid CFL EoS constructed, considering two distinct cases: a) B = 140 MeV·fm$^{-3}$ and b) Δ = 40 MeV, while varying the other parameter in each scenario. It is important to note that the selected B-Δ combinations fall outside the stability window shown in Fig. 1. This is critical, as non-absolutely stable states are required; otherwise, pure CFL matter would be energetically favored over the hybrid configuration. The results indicate that the models can reproduce the HESS J1731-347 event via a hybrid branch, which initiates between 0.5 and 1$M_\odot$. However, the phase transition introduces a softening in the EoS, causing the maximum masses to fall well below 2$M_\odot$. This discrepancy suggests that while the models can account for the HESS J1731-347 event, they are not consistent with other measurements, necessitating further refinement and alternative approaches. Lastly, the different behavior of quark and hybrid stars at low masses lies in the fact that the latter are surrounded by a crust with a specific equation of state, which is significantly different from a quark equation of state [59]. As a result, at low masses, hybrid stars are characterized by an extended crust, leading to larger radii for the same mass values.

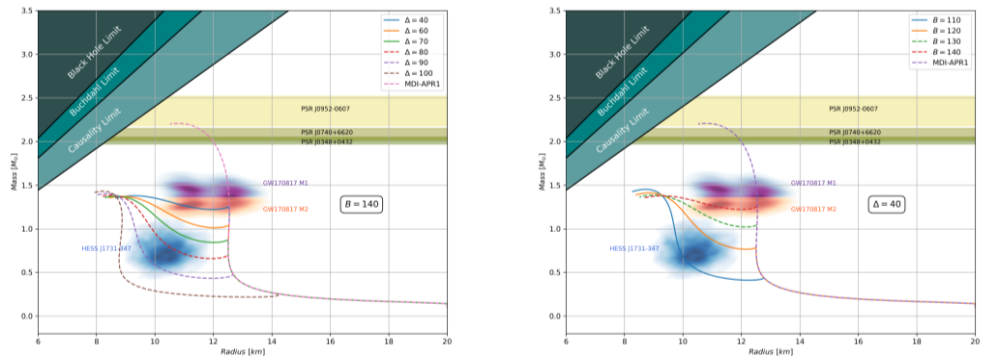

**Figure 3**. *M-R diagrams for the constructed hybrid models are presented for varying Δ and a fixed B=140 (left panel) and for varying B and a fixed Δ=40 (right panel) (B in MeV · fm$^{-3}$, Δ in MeV).*

## CONCLUSIONS

In this study, we conducted a comprehensive analysis of the CFL quark matter EoS, constraining its parameters to align with recent measurements of the HESS J1731-347 CCO, as well as the GW170817 event and the heaviest known pulsar, PSR J0952-0607. Our results suggest that CFL quark matter can explain all the previously mentioned measurements, within a parameter space shown in Fig. 2 (right). Within this parameter window, pure CFL matter satisfies all observational criteria, while also respecting causality constraints. Therefore, CFL quark matter could be a promising candidate for the HESS J1731-347 CCO. However, in the case of the hybrid CFL models developed, we find that although a hybrid branch can indeed reproduce the extremely low mass and radius HESS J1731-347 event, the maximum masses predicted by this model fall significantly below those observed for the most massive pulsars. A potential solution to this issue could involve incorporating a density-dependent bag parameter, which decreases at higher densities, thereby stiffening the EoS, as suggested in Ref. [43]. While the exact nature of this object remains uncertain, ongoing research, coupled with advances in observational techniques, will be pivotal in resolving this enigmatic problem.

### Acknowledgments

K. K. would like to thank Mr. Stefanos Kargas for his valuable assistance with the computational part of this study. (P.L.-P.) The research work was supported by the Hellenic Foundation for Research and Innovation (HFRI) under the 5th Call for HFRI PhD Fellowships (Fellowship Number: 19175).